\documentstyle[preprint,aps,prd,psfig]{revtex}
\begin{document}
\renewcommand{\thesection}{\arabic{section}}
\renewcommand{\thesubsection}{\arabic{subsection}}
\topskip 1.5cm
\title{Pauli Paramagnetism of Neutron Star Matter and the Upper Limit
for Neutron Star Magnetic Fields}

\author{
 Soma Mandal$^{a)}$ {\thanks{E-Mail: soma@klyuniv.ernet.in}}
 and Somenath Chakrabarty$^{a),b)}${\thanks{E-Mail: 
 somenath@klyuniv.ernet.in}} }
\address{
$^{a)}$Department of Physics, University of Kalyani, Kalyani 741 235,
India and
$^{b)}$Inter-University Centre for Astronomy and Astrophysics, Post Bag 4,
Ganeshkhind, Pune 411 007, India
}
\date{\today}
\maketitle
	\begin{abstract}
A relativistic version of Pauli paramagnetism for $n-p-e$ system inside 
a strongly magnetized neutron star has been developed. An analytical 
expressions for the saturation value of magnetic field strength for each of 
these constituents at which they are completely polarized have been obtained.
From the fully polarized configuration of electronic component, an upper
limit for neutron star magnetic field is predicted. It has been concluded that
indeed, magnetars, as stronly magnetized young neutron stars can not
exist if the constituents are electron, proton and neutron in
$\beta$-equilibrium. An alternative model has been proposed. 
	\end{abstract}

\section{Introduction}
	
The study of the effect of strong quantizing magnetic field of neutron stars 
on dense nuclear matter has gotten a new dimension after the discovery of a 
few magnetars. These exotic objects are believed to be strongly magnetized 
young neutron stars of surface magnetic field $\approx 10^{15}$G 
\cite{R1,R2,R3,R4,R5}. The use of scalar virial theorem shows that the 
magnetic field strength at the core region may go upto $10^{18}$G \cite{R6}. 
It is therefore very much advisable to study the effect of such strong magnetic
field on various physical properties of dense neutron star matter as well as 
on various physical processes taking place inside neutron stars.  An extensive 
studies have already been done on the equation of state of dense neutron star 
matter in presence of strong magnetic field \cite{R7,R8,R9,R10}.  Such studies 
are based on quantum mechanical effect of strong magnetic field. The effect of 
strong quantizing magnetic field on the gross properties, e.g., the mass, 
radius, moment of inertia etc., of neutron stars, which are strongly dependent 
on the equation of state of matter have also been obtained \cite{R11}. In the 
second kind of studies, how the weak processes (reactions and decays) are 
affected by the quantum mechanical effect of strong magnetic field have been 
obtained \cite{R13,R14}. As a consequence the $\beta$-equilibrium condition 
also depends on the strength of magnetic field. Since the cooling of neutron 
stars is dominated by the emission of neutrinos produced by weak processes 
inside the stars, these studies also give an idea of the effect of strong 
magnetic field on thermal evolution of neutron stars. Not only that, the
presence of strong magnetic field can change significantly, both qualitatively 
and quantitatively the transport coefficients (viscosity, thermal conductivity,
electrical conductivity etc.) of dense neutron star matter \cite{R15,R16,R17}. 
The magnetic field can change the tensorial character of transport coefficients
of neutron star matter in presence of strong magnetic fields. Such qualitative
change in transport coefficients can cause some significant changes in thermal 
evolution of neutron star matter and the evolution of its magnetic field. 
There are another kind of studies; the effect of strong quantizing magnetic 
field on quark-hadron phase transition \cite{R18,R19}. It was shown explicitly 
that a first order quark-hadron phase transition is absolutely forbidden if the
strength of magnetic field exceeds $10^{15}$G. However, a metal insulator type 
(color insulator to color metal) second order phase transition is possible 
unless the field strength exceeds $10^{20}$G. It has also been shown, that 
even if there is a first order quark-hadron phase transition for magnetic field
strength  $< 10^{15}$G at the core region of a neutron star, an investigation 
of chemical evolution of quark matter, with various initial conditions leads to
the system in $\beta$-equilibrium, revealed that the system becomes 
energetically unstable in chemical equilibrium \cite{R19}.  In some completely 
different type of studies, the stability and some of the gross properties of 
deformed stellar objects are analyzed with general relativity 
\cite{R20,R21,R22,R23,R24}. The presence of strong magnetic field destroys the
spherical symmetry of neutron star. Then it is possible for a deformed and 
rotating neutron star to emit gravity waves, which in principle may be 
detected.  Very recently we have critically studied the ferro-magnetism of 
neutron star matter which could be one of the sources of residual magnetism of 
old neutron stars/sources  of magnetic field of millisecond pulsars. In these 
studies, we have shown that a spontaneous Ferromagnetic transition in absence 
of external magnetic field is not possible in neutron star matter in 
$\beta$-equilibrium. However in the case of neutrino trapped neutron star 
matter (Proto-neutron star matter), the possibility  of such a transition can 
not be ruled out, provided the neutrinos carry some non-zero mass \cite{R25}.
We have also analyzed the problem with the occupancy of zeroth Landau levels 
by electrons/protons, which occur in presence of ultra-strong magnetic fields 
\cite{R26}. It has been argued in this critical analysis that in presence of 
a strong quantizing magnetic field the existence of neutron star matter in 
$\beta$-equilibrium is questionable. Which further opens up a vital question 
on the possibility of magnetars as young and strongly magnetized neutron stars.
	
In this paper we shall present a relativistic version of Pauli paramagnetism of
neutron star matter in $\beta$-equilibrium. We shall extend the work of 
Shul'man \cite{R27} to the relativistic region (i.e., in the high density 
regime of neutron star matter) and study the paramagnetism of relativistic
nuclear matter. The aim of this paper is to show that the fully polarized 
configuration of electronic components puts a restriction on the upper limit 
of neutron star magnetic field.  We have also discussed the alternative picture
of neutron star structure, if the magnetic field exceeds that limit.  The paper
is organized in the following manner. In section 2, we shall develop the 
formalism of relativistic version of Pauli paramagnetism, in section 3 we 
study the Pauli paramagnetism of neutron star matter and in the last section, 
we shall conclude and discuss the importance of this work.
\section{Formalism}
We consider a Fermion gas of density $n$ in presence of a strong magnetic 
field $B$ at zero temperature ($T=0$). Then the number density is given by 
\cite{R27}
\begin{equation}
n=\frac{[2\mu B(2\mu B+2m)]^{3/2}(2x^{3/2}-1)}{6\pi^2}
\end{equation}
where $\mu$  and $m$ are respectively the magnetic dipole moment and mass of 
the particle and
\begin{equation}
x=\frac{\epsilon^2(B)-m^2}{2\mu B(2\mu B+2m)}
\end{equation}
is a dimensionless quantity, known as the energy variable and
\begin{equation}
\epsilon (B)=[m^2+x[R_{3/2}(x)]^{-2/3}(\epsilon_0^2-m^2)]^{1/2}
\end{equation} 
is the single particle energy, $\epsilon_0$  is the value of $\epsilon(B)$  
for $B=0$ and
\begin{equation}
R_{3/2}(x)=\frac{1}{2}(2x^{3/2}-1)
\end{equation}
Then it is a matter of simple algebraic manipulation to obtain the number 
densities with spin parallel and anti-parallel to the direction of magnetic 
field $B$, and are given by
\begin{eqnarray}
n^\uparrow(B)&=&\frac{n}{2}x^{3/2}[R_{3/2}(x)]^{-1} \nonumber \\
n^\downarrow(B)&=&n^\uparrow (B)(1-x^{-3/2})
\end{eqnarray}
It is obvious from these two expressions that for $x\rightarrow \infty$, for 
which $B\rightarrow 0$ (no field situation), $n^\uparrow=n^\downarrow=n/2$,
which indicates that both up and down spin states are equi-probable. On the 
other hand for $x\rightarrow 1$ we have $n^\uparrow=n$ and $n^\downarrow=0$, 
which are the saturation densities (fully polarized scenario). In the case of 
spin saturation, the single particle energy is given by
\begin{equation}
\epsilon(B)=\epsilon^{(s)}(B)=[2^{2/3}\epsilon_0^2-(2^{2/3}-1)m^2]^{1/2}
\end{equation}
Since for $x<1$, the value of $n^\downarrow$ becomes negative, the lowest 
value of $x$ is $1$ and the corresponding saturated value for magnetic field 
strength is given by
\begin{equation}
B_s=\frac{1}{2\mu}[\{(6\pi^2n)^{2/3}+m^2\}^{1/2}-m]
\end{equation}
This is also the maximum value of magnetic field strength which the neutron 
star matter can sustain.  In fig.(1) we have plotted the values of 
$n^\uparrow$(B) and $n^\downarrow$(B) (in terms of $n$) for a Fermi system 
against $x$. The upper curve is for $n^\uparrow$(B) and lower one corresponds 
to $n^\downarrow$(B). It is obvious from the figure that $x$=$1$  corresponds 
to fully polarized scenario, whereas $x\rightarrow\infty$  corresponds to 
completely unpolarized picture.
\section{Pauli Paramagnetism of Nuclear Matter}
We next consider a $n-p-e$  system inside a neutron star in presence of a 
strong magnetic field at $T$ = $0$ (since the chemical potential $\mu_i*$ for 
the $i$th species is $>>$ the temperature of the system $T$). Now in presence 
of a strong magnetic field, the chemical potential for the $i$th component is 
given by
\begin{equation}
\mu_i^*=\epsilon_i(B)-\mu_iB
\end{equation}
where $i=n, p$, or $e$. The charge neutrality condition gives $n_p$ = $n_e$ 
where $n_i$  is the number density of $i$th species.  Finally, the baryon 
number density is given by
\begin{equation}
n_B=n_n+n_p
\end{equation}
which we treat as a constant parameter. It has been noticed from some critical 
investigations that the $\beta$-equilibrium condition can not be achieved in a
polarized neutron star matter ($n-p-e$  system in presence of a strong magnetic
field \cite{R26}). We have assumed  some arbitrary iso-spin symmetry, defined 
by the parameter
\begin{equation}
\beta=\frac{n_n-n_p}{n_n}
\end{equation}
 with $0\le\beta\le1$ (since the matter with excess protons is highly unstable,
 we have not considered negative values of $\beta$). Then combining eqns.(9) 
 and (10) we have
\begin{equation}
n_n=\frac{n_B}{2-\beta}{\rm{~~~~and~~~~}}
n_p=n_n(1-\beta)
\end{equation}
Hence after a little algebraic manipulation, we have from eq,(2), the
dimensionless energy variable for the species $i$ 
\begin{equation}
x_i=\left [0.5+\left \{\frac{2\mu_i B(2\mu_i
B+m_i)}{(3\pi^2n_i)^{2/3}}\right \}^{-2/3}\right ]^{3/2}
\end{equation}
It clearly shows that the dimensionless energy variables for the
constituents are functions of magnetic field strength B, density of matter
$n_B$ and the iso-spin symmetry parameter $\beta$. We have taken
$\mu_p=2.79\mu_N$, $\mu_n=-1.91\mu_N$ and $\mu_e=1.001\mu_B$, where
$\mu_N=3.15\times10^{-14}MeV/T$ and $\mu_B=5.79\times10^{-11}MeV/T$ are
respectively the nuclear magneton and Bohr magneton. In figs.(2), (3)
and (4) we have shown that the variations of $x_i$ ($i=n, p, e$) with $B$
for $n_B=2n_0$ and $\beta=0$, $0.5$ $1.0$ (to avoid some numerical
problem we have taken $x=0.99$ instead of $1.0$). All these figures show
that within the range of magnetic field strength relevance for
magnetized neutron stars, neutrons and protons never become fully
polarized ($x_n, x_p >>1$) whereas the value of $x_e$ is very close to
unity which means that the electrons are in very close to fully polarized
configuration. In figs.(5), (6) and (7) we have plotted the same 
quantities for $n_B=6n_0$. In these sets of figure also very close to
fully polarized states for electrons is possible whereas neutrons and
protons remain in the states with almost unpolarized configuration.
However both the neutron matter and the proton matter can be made fully
polarized by increasing the strength of magnetic field. 
\section{conclusions}
We have noticed that 
although the electronic component becomes fully polarized for $B\sim 10^{16}$G, 
the nuclear matter can only become fully polarized if and only if the 
magnetic field strength exceeds $10^{20}$G. We have also observed that for such 
strong magnetic fields make $n^\downarrow_e$ negative, which is completely
unphysical.  Therefore, the fully polarized configuration of electronic
component restricts the upper limit of a neutron star magnetic field to
$\approx 10^{16}$G. However the inclusion of pions $(\pi^-)$ or
kaons $(K^-)$ instead of electrons does not impose any restriction on the
upper limit. At the same time the inclusion of these two components
instead on electrons allow the $n-p$ system to become fully polarized even in
$\beta$-equilibrium condition. We can therefore conclude that the
magnetars, if they exist at all and are assumed to be strongly
magnetized young neutron stars, the constituents are possibly
$n-p-\pi^-$ or $n-p-K^-$ instead of $n-p-e$. Otherwise the existence of
such objects is impossible. The incorporation of interaction in
nuclear matter sector does not change the conclusion. The observed maximum
polarization in the electronic sector whereas almost unpolarized
configuration of nuclear matter regime is because of several orders of 
magnitude difference in the magnitudes of magnetic dipole moments. In the case 
of electron it is $\sim$ Bohr magneton, whereas in the case of nucleons it is 
$\sim$ nuclear magneton. This is independent of the type of interaction in the
neutron star matter.
%----------------- Figures --------------------------
\begin{figure} 
\psfig{figure=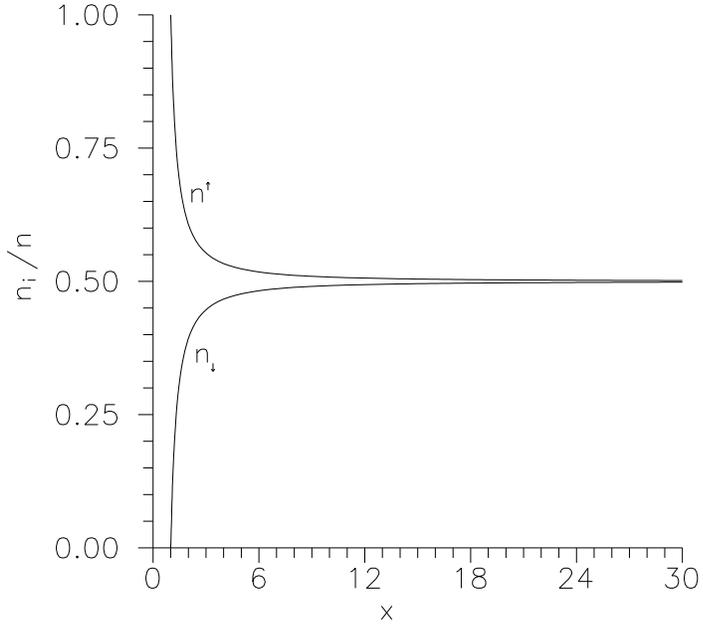,height=0.5\linewidth}
\caption{
The variation of $n^\uparrow$ and $n^\downarrow$ (in terms of $n$) with
$x$.
}
\end{figure}
\begin{figure} 
\psfig{figure=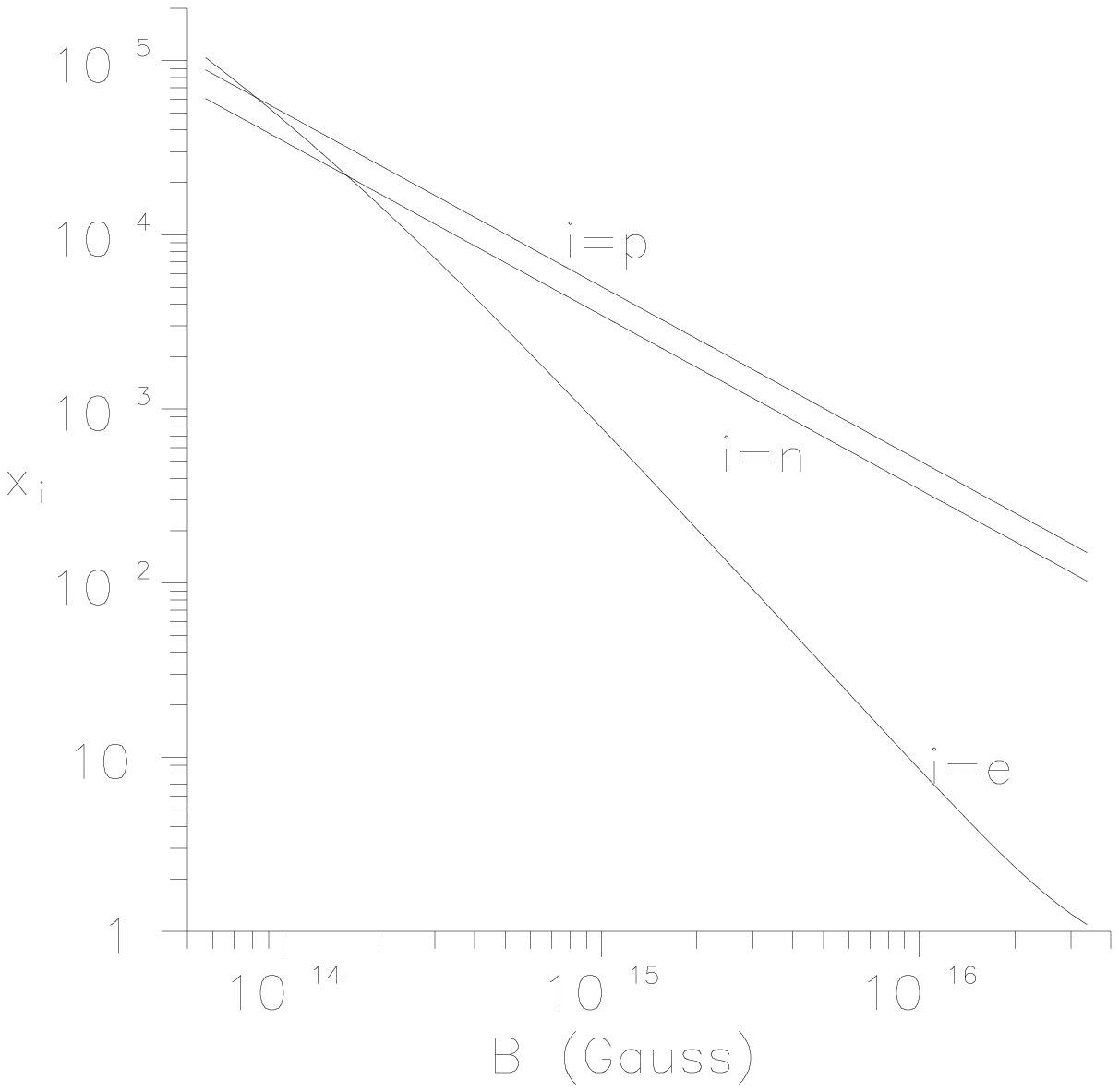,height=0.5\linewidth}
\caption{
Variation of dimensionless energy variables $x_i$'s with $B$, expressed
in terms of $B_c^{(e)}$, for $n_B=2n_0$ and $\beta=0$.
}
\end{figure}
\begin{figure} 
\psfig{figure=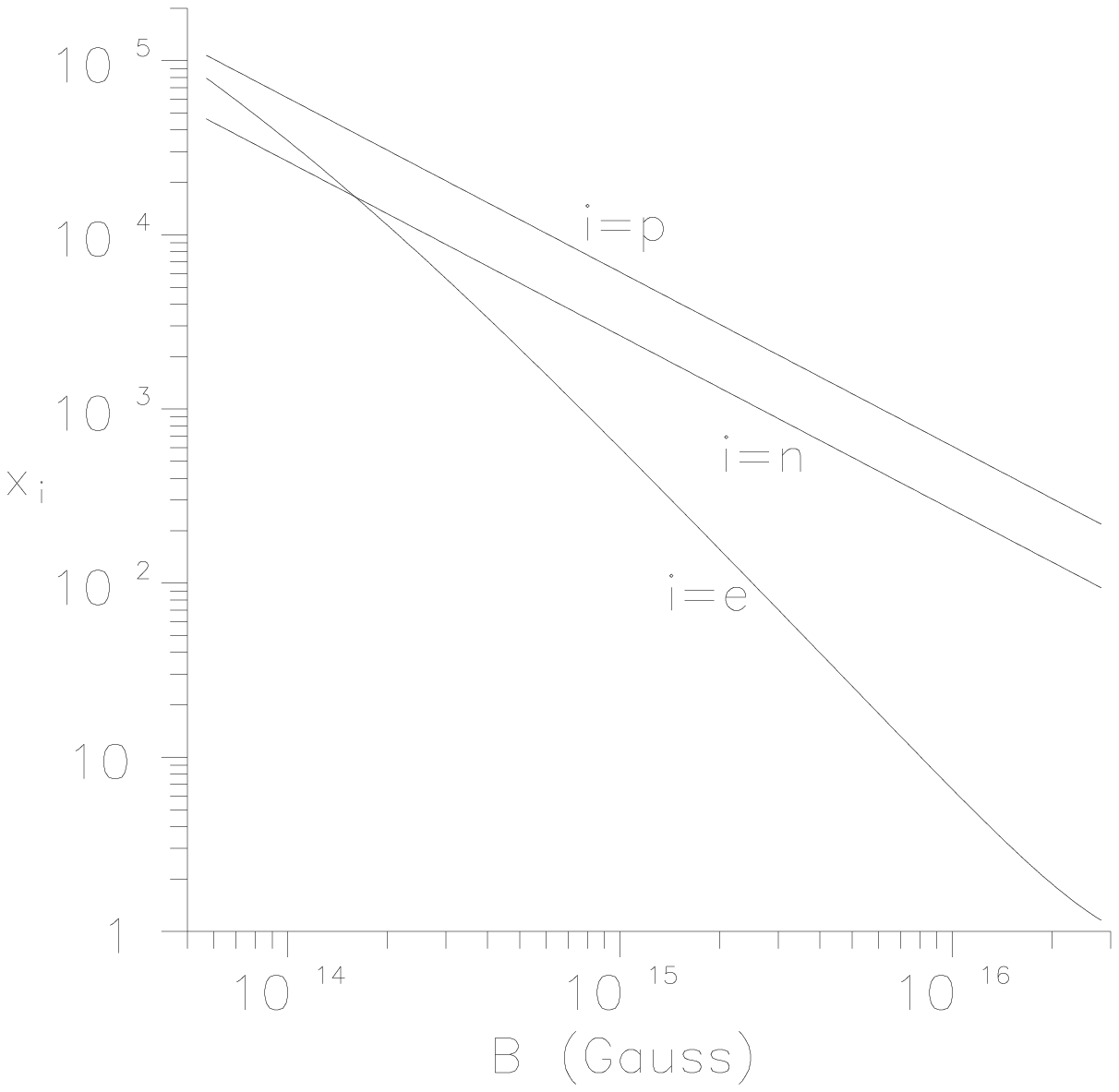,height=0.5\linewidth}
\caption{
Variation of dimensionless energy variables $x_i$'s with $B$, expressed
in terms of $B_c^{(e)}$, for $n_B=2n_0$ and $\beta=0.5$.
}
\end{figure}
\begin{figure} 
\psfig{figure=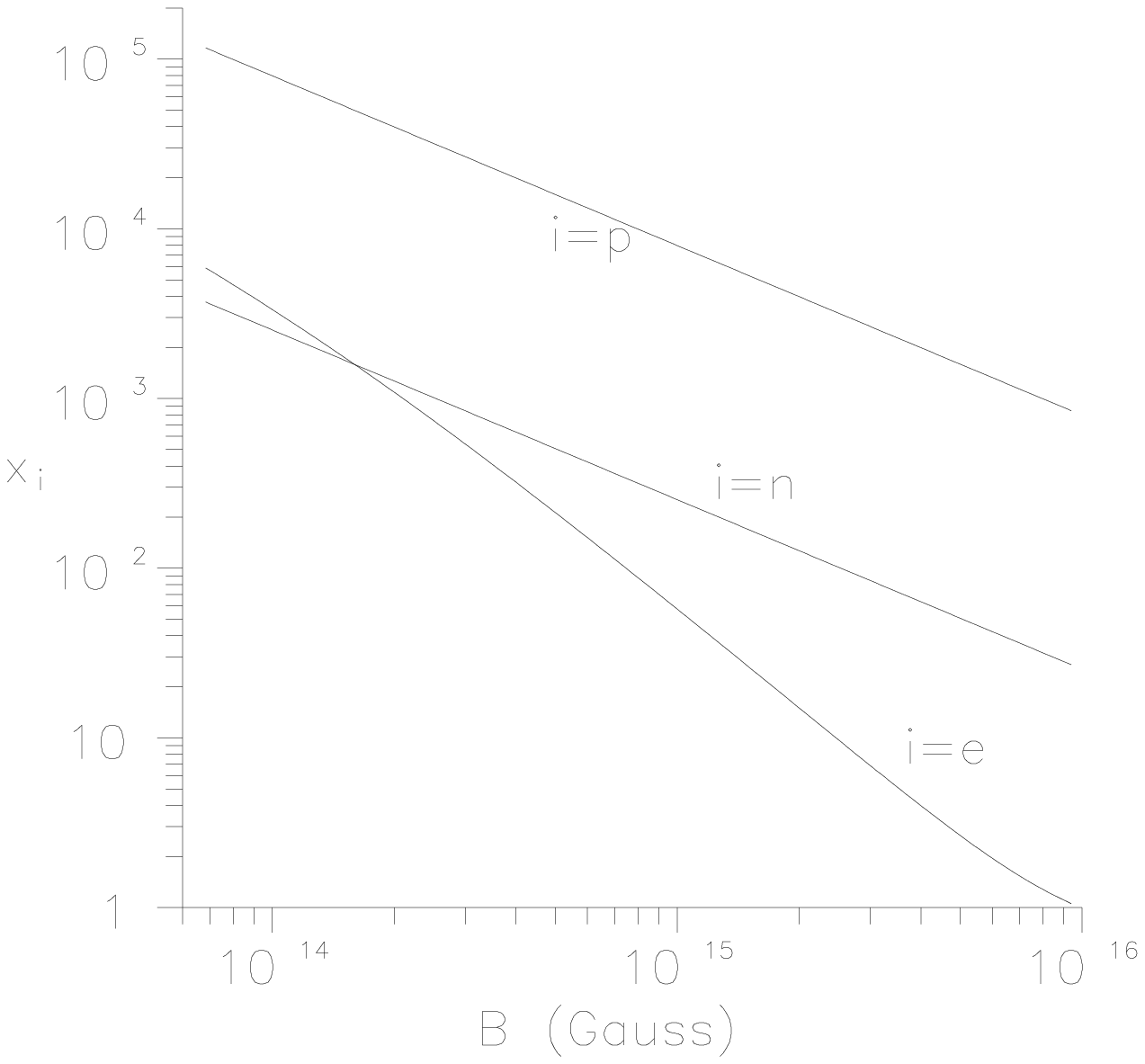,height=0.5\linewidth}
\caption{
Variation of dimensionless energy variables $x_i$'s with $B$, expressed
in terms of $B_c^{(e)}$, for $n_B=2n_0$ and $\beta=0.99$.
}
\end{figure}
\begin{figure} 
\psfig{figure=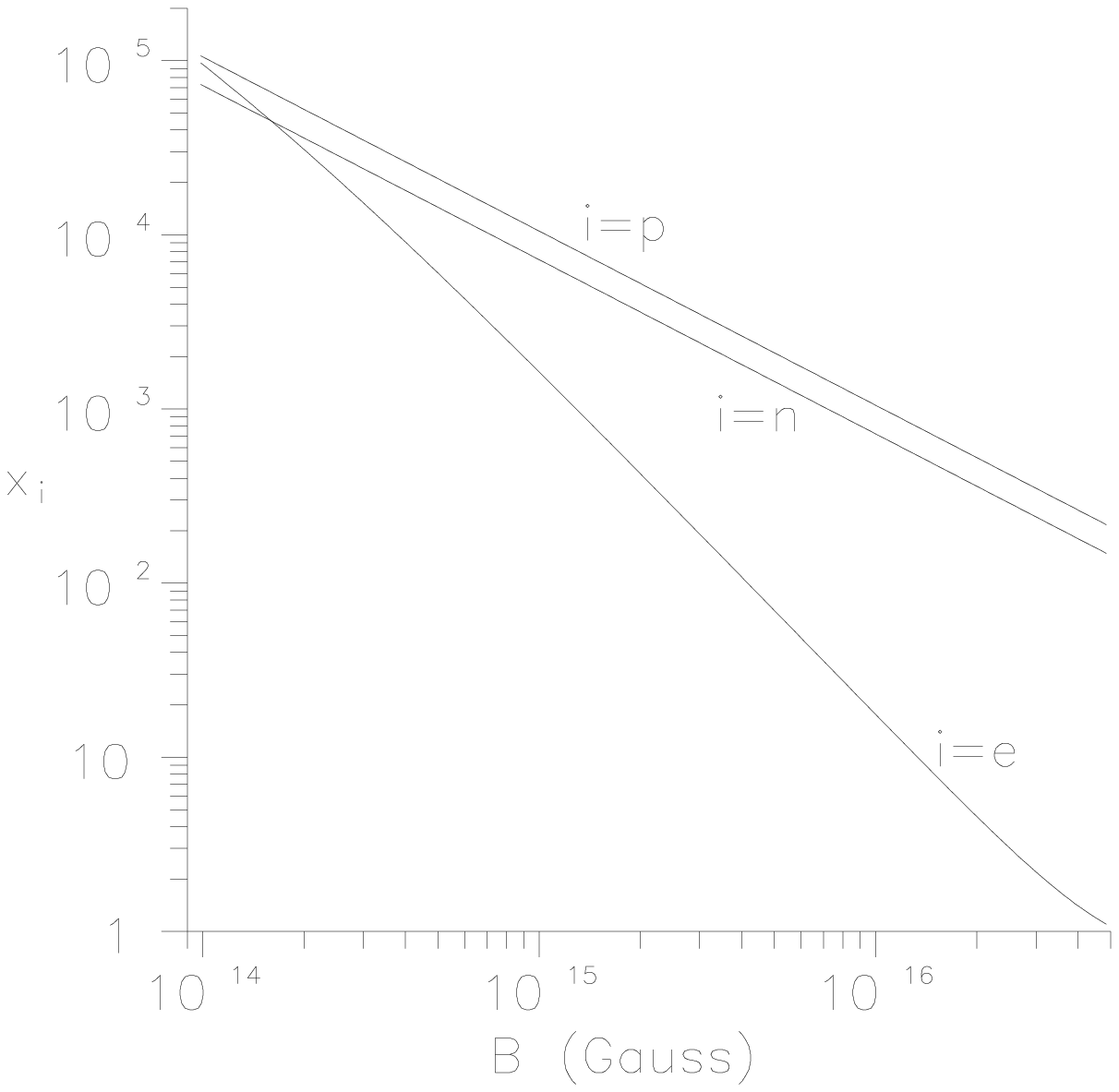,height=0.5\linewidth}
\caption{
Variation of dimensionless energy variables $x_i$'s with $B$, expressed
in terms of $B_c^{(e)}$, for $n_B=6n_0$ and $\beta=0$.
}
\end{figure}
\begin{figure} 
\psfig{figure=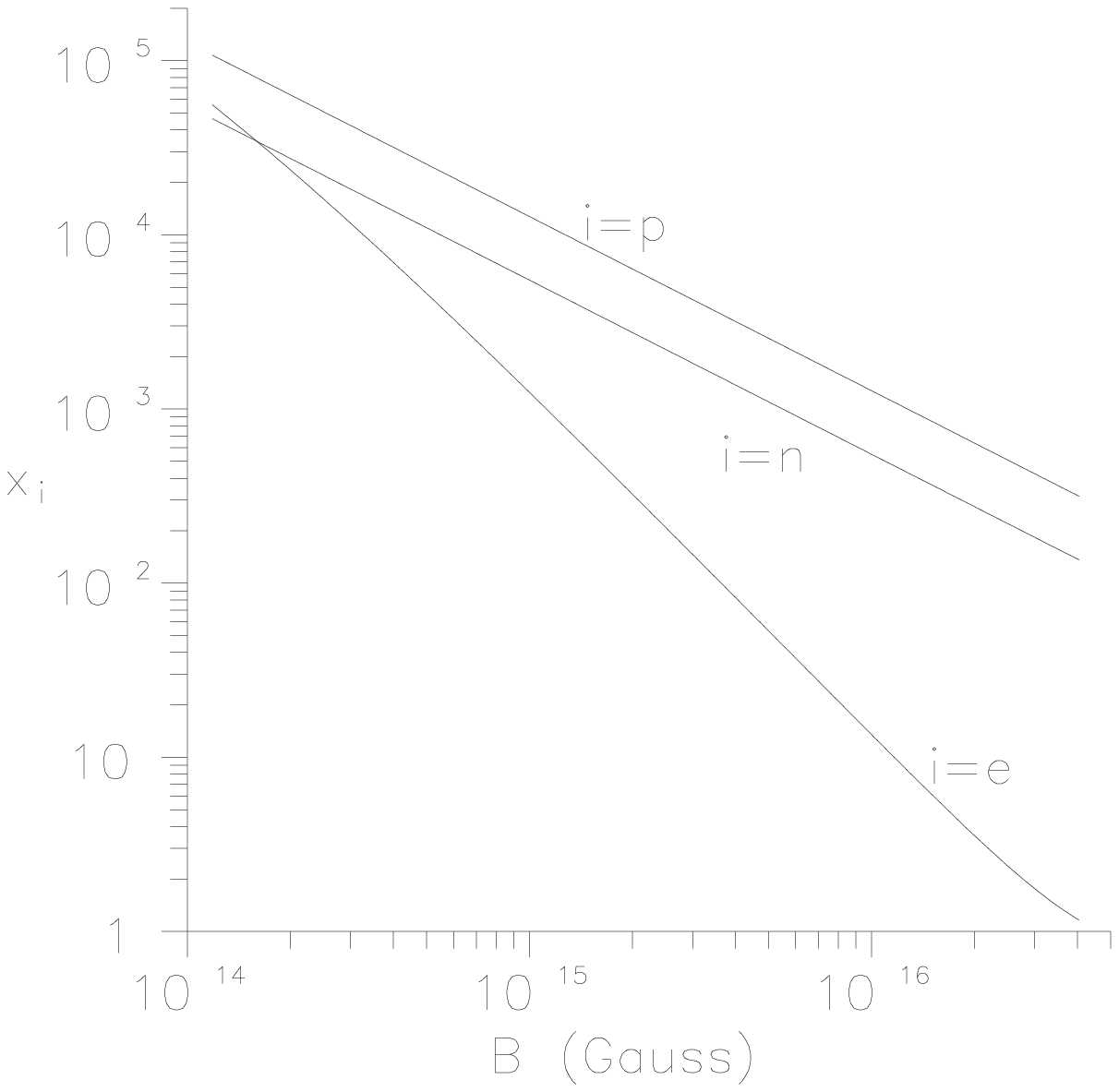,height=0.5\linewidth}
\caption{
Variation of dimensionless energy variables $x_i$'s with $B$, expressed
in terms of $B_c^{(e)}$, for $n_B=6n_0$ and $\beta=0.5$.
}
\end{figure}
\begin{figure} 
\psfig{figure=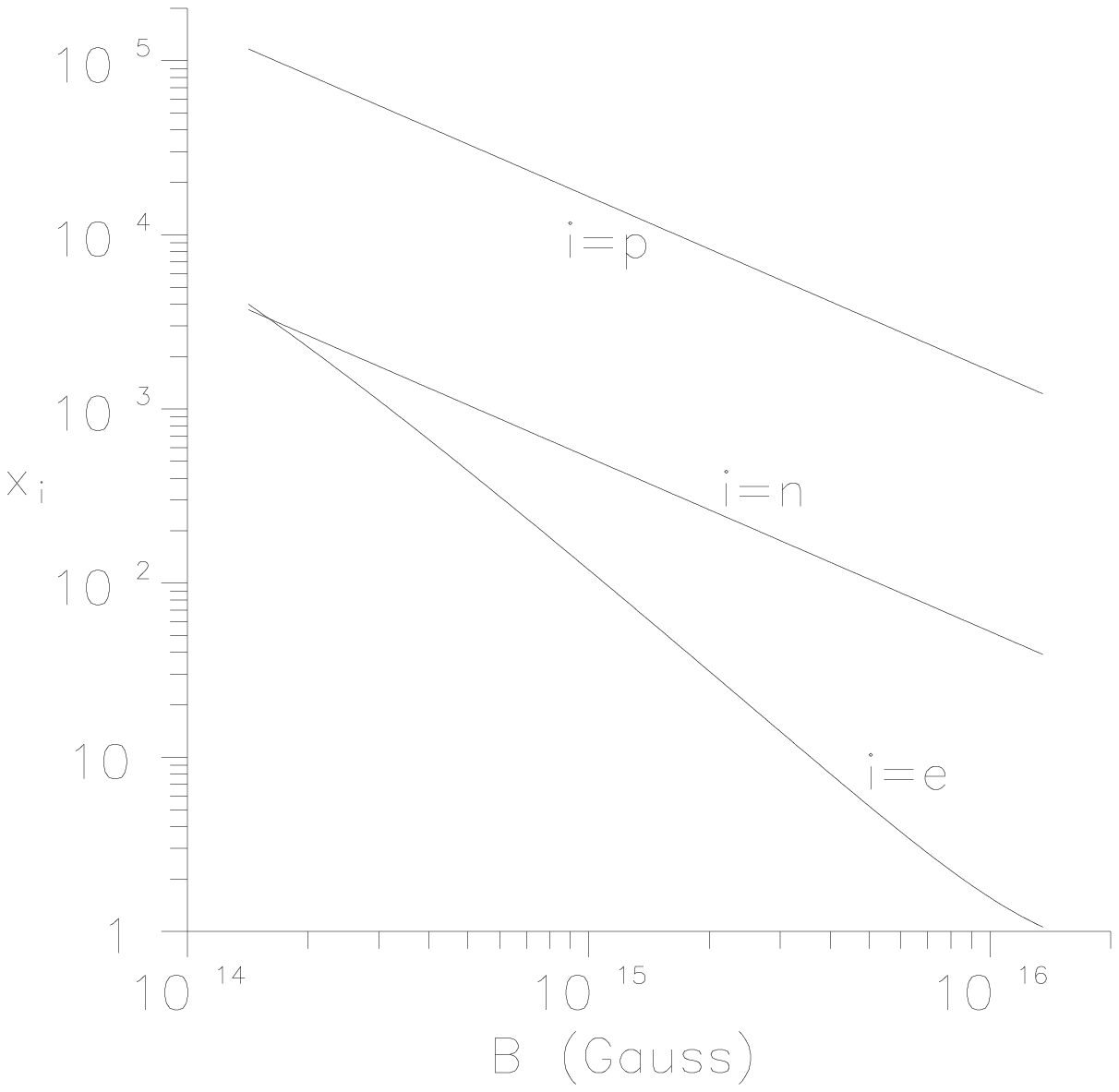,height=0.5\linewidth}
\caption{
Variation of dimensionless energy variables $x_i$'s with $B$, expressed
in terms of $B_c^{(e)}$, for $n_B=6n_0$ and $\beta=0.99$.
}
\end{figure}
%-------------- References --------------------------

\end{document}